\documentstyle[proceedings,bibstyle]{crckapb}

\def\spose#1{\hbox to 0pt{#1\hss}}
\def\lta{\mathrel{\spose{\lower 3pt\hbox{$\mathchar"218$}}
     \raise 2.0pt\hbox{$\mathchar"13C$}}}

\begin{opening}
\title{FUNDAMENTAL ASPECTS OF THE ISM FRACTALITY}
\author{D. PFENNIGER}
\institute{Geneva Observatory, University of Geneva\\
           CH-1290 Sauverny, Switzerland}
\end{opening}

\runningtitle{ISM FRACTALITY}

\begin{document}

\begin{abstract}
The ubiquitous clumpy state of the ISM raises a fundamental and open
problem of physics, which is the correct statistical treatment of
systems dominated by long range interactions.  A simple solvable
hierarchical model is presented which explains why systems dominated
by gravity prefer to adopt a fractal dimension around 2 or less, like
the cold ISM and large scale structures.  This has direct relation
with the general transparency, or blackness, of the Universe.
\end{abstract}

\section{Introduction: Clumpiness in Astrophysical Gases}
The extreme inhomogeneity, the supersonic turbulence, and multi-phase
nature of the interstellar medium (ISM), meaning an extreme departure
from thermodynamical equilibrium, was well recognised many decades ago
(e.g. von Weizs\"acker 1951).  Over the years, new wavelengths and
higher angular resolution have revealed several times that the
interstellar gas was more clumpy than accepted before.  Often, it was
argued that the smallest clouds had been resolved, to be disproved
later on (typically they span ``three times the beamwidth'', Verschuur
1993).

Clearly, the radically opposite behaviours of cosmic and terrestrial
gases must be understood.  Since the ISM covers many decades of
density and temperature, very different physical processes, such as
radiation transfer, chemistry, magnetic fields, and gravity are
simultaneously involved, confusing the issue about the relative
importance of each one.  Star formation can even be viewed as an
extreme prolongation of the ISM inhomogeneity.  The concurrent
interplay of so many factors makes the study of the ISM formidable.
Moreover, different kinds of devices and skills are needed to unveil
its fullness, increasing the communication problem across specialties.

With respect to terrestrial gas, the new factor in the ISM and larger
scale structures is, of course, self-gravity.  Although there are
discussions whether the molecular clouds are bounds by gravity or just
marginally so (e.g. Blitz 1993), in any case gravity pervades cosmic
structures from the largest galaxy superclusters down to stars and
planets.  By continuity its r\^ole must be expected to be generally
dominant throughout the scales.

\section{Homogeneity in the Classical Gas}
Before discussing the inhomogeneity of cosmic gas, one should clearly
understand why in terrestrial conditions gas tends toward homogeneity.
In terrestrial situations a classical gas is {\it confined \/} by
exterior factors (walls, etc.), which means that in the Lagrangian of
the $N$ gas particles,
\begin{equation}
 L(q_i, \dot q_i) = T(\dot q_i) - U(q_i),\quad\mathrm{and}\quad U(q_i)/T(\dot q_i) \to 0,
\quad (i=1,\ldots N),
\end{equation}
the kinetic energy term $T(\dot q_i)$ dominates the interaction term
$U(q_i)$.  The motion of individual particles being highly chaotic,
sensitive to perturbations, no integrals beside the classical
integrals does exist, and the system can be assumed ergodic in good
approximation.  Typically, perturbations (``the walls'') reshuffle the
particles in phase space while preserving in average the scalar energy
integral.  This allows the average time invariance of the system.  But
the global {\it vector\/} integrals (angular momentum, centers of
velocity and mass) must vanish in the reference frame of the box.
Since the system (1) is locally dominated by the kinetic term, it must
be translationally invariant in space.  As well known to this symmetry
corresponds spatial homogeneity.  At scales much larger than the
interparticle distances the gas must tend to be smooth,
i.e.~differentiable.  The use of hydrodynamics and other differential
equations is then justified.

In such a situation the total energy of the system is proportional to
the volume, i.e. extensive.  In order to be able to take the large $N$
limit, the extensivity of energy is an additional requisite in usual
statistical mechanics.

Thus, the a priori complex behaviour of a gas can be statistically
simple when a global symmetry, such as translation invariance, exists.
However, when the interaction potential becomes non-negligible at
lower temperature this symmetry may be broken.  Often a phase
transition reveals then another symmetry (e.g. when crystallisation
occurs).  But if the interactions are short range, then energy is 
still extensive in good approximation.

In contrast, when the interactions are long range and attractive, as
for gravity, they can never be neglected in the large $N$ limit, and
energy is no longer extensive: the ground to expect a statistical
homogeneous state, even locally, is lost.  Curiously, the
inapplicability of thermodynamics to gravitating systems is well known
by thermodynamicists (e.g.~Jaynes 1957; Prigogine 1962), but often
ignored in astrophysics.  

The association of thermodynamics with gravity leads immediately to
paradoxes, such as the occurrence of negative specific heats
(Lynden-Bell \& Lynden-Bell 1977), implying that a thermal equilibrium
is impossible!

The ideal model of the gravitating perfect gas enclosed in a spherical
box at a fixed temperature shows itself the limit of the association
(see Binney \& Tremaine 1987, Fig.~[8-1]).  The isothermal sphere
equilibrium does exist only above a critical temperature, which
corresponds to the Jeans' critical temperature.  At sufficiently high
temperature the system energy is positive and a box is required to
confine the system, as for usual terrestrial gas.  At some lower
temperature the system energy is negative, so is bound by gravity, but
a thermal equilibrium still exists: this is the regime of the stars.
However, below the Jeans' critical temperature, no equilibrium
solution does exist which fulfills the local homogeneity assumption.
The self-gravity then dominates.  The open problem is then to
characterise the asymptotic statistical state of the perfect gas {\it
below the Jeans' critical temperature}.  Clearly, this concerns many
astrophysical systems, from most of the ISM to the large-scale
cosmological structures.

\section{Statistical Equilibrium with Long-Range Interactions}
Before introducing any additional physics relevant to the ISM, the
statistical behaviour resulting just from the interplay between
gravity and dynamics should be characterised.  This partial problem is
a highly idealised simplification of the full complexity of the ISM,
but no progress can be expected if this fundamental aspect remains
obscure.

The classical gas is based on a very rough model of particles without
interaction.  Yet, more important it includes the symmetry of
translational invariance.  In the ISM and large scale structures, the
dominant symmetry is {\it scale-invariance\/}.  The Lagrangian of the
gas particles is
\begin{equation}
 L(q_i, \dot q_i) = T(\dot q_i) - U(q_i),\quad \mathrm{with} \quad
U(\alpha q_i) = \alpha^p U(q_i), \quad p=-1.
\end{equation}
The idea is to represent this empirical fact by a hierarchical system
{\it statistically scale-invariant}.  In many cases of solid state
physics, systems in phase transition build also long-range
correlations with universal scaling laws; interestingly, even
surprisingly crude models, such as the Ising model, reproduce
remarkably well the measured scaling laws (e.g.~Stanley 1995). 
Important is therefore to include the right symmetry in the model.

Here we suppose that the system is made of hierarchical mass {\it
clumps\/}, each clump being made, in average, of $N$ sub-clumps,
recursively over a number $L$ of levels above the ground level 0
(see Pfenniger \& Combes 1994, PC94). A clump is characterised by a
finite mass $M$ and a finite length scale $r$.  The mass distribution,
to be scale-invariant, scales as a fixed power $D$, which defines the
mass fractal dimension.  Thus,
\begin{equation}
M_L = N M_{L-1} = N^L M_0,  \qquad \mathrm{ and} \qquad
M_L = M_0 \left(r_L/r_0\right)^D.
\end{equation}
In real physical systems a lower and upper levels must exist, which
define, analogously to walls in the classical gas, the boundary
conditions.  The boundary conditions are convenient to abstract the
complications coming from the external world.  For example: 1) the
lowest level of fragmentation is reached in the ISM when the heat
transfer time exceeds the dynamical time (see PC94); 2) the upper
level in cosmological structures is given by the time-dependence
introduced at the largest scale by the universal expansion, which
proceeds at a slower pace than the small-scale clustering.

The second hypothesis concerns the particle interactions.
Generalising the gravitating case, the potential $\Phi$ is supposed to
be a power law, i.e. $\Phi = G M r^p/p$ ($p=-1$ for gravity, and $G>0$
for attracting interactions).  Suggested by the system hierarchical
organisation, we approximate the potential energy $U_L$ at level $L$
by
\begin{equation} 
U_L = N U_{L-1} +
\frac{G}{p} M_L^2 r_L^p \left( 1 + \alpha \frac{r_{L-1}}{r_L}\right) .
\end{equation}
The term with $\alpha$ represents tidal interactions.  Since the scale
ratio is constant, this term is constant and can be absorbed in a new
coupling constant $G'$.  We will see that the main results do not
depend on the {\it value\/} of $G$ (indeed scale invariance effects
depend on the power law exponent $p$).

For parameters suited to the ISM ($D\approx 1-2$, $N=5-8$, Scalo 1990)
the approximation (4) has been checked to be accurate at the percent
level.
 
The third hypothesis is of statistical character.  Thermodynamics can
not be used since it excludes long-range interactions.  The only
remaining statistical tool is the virial theorem.  If a scale
invariant system finds a statistical equilibrium, at each level the
virial theorem must hold,
\begin{equation} 
2T_L - p U_L = 3 P_{L+1} V_L, 
\end{equation}
where $T_L$ is the total kinetic energy cumulated up level $L$,
$P_{L+1}$ is the outer pressure, and $V_L \sim r_L^3$ the clump
volume.  Here the outer pressure is purely kinetic: it is given by
$2/3$ of the kinetic energy density outside the sub-clumps.
Approximately,
\begin{equation}
P_L = \frac{2}{3} \frac{T_L - N T_{L-1}}{V_L - N V_{L-1}}, 
\end{equation}
which neglects the superpositions of clumps during collisions.  For
the sake of simplicity, in PC94 the pressure term was neglected, yet
the clump ``collisions'' were determined to be frequent for the
typical parameters of the ISM.  Clump collisions may well disrupt or
merge them, but the supposed Jeans unstable medium (due e.g. to fast
cooling) also fragments clouds fast, reforming the clumps before a
crossing time.  Therefore, in a statistical equilibrium the average $N$
should be constant, and at any moment the fraction of colliding clumps
should be small.

The above recurrences for $M_L$, $V_L$, $U_L$ and $T_L$ in Eqs. [3-6]
can be solved exactly in finite terms. With $x\equiv r_L/r_{L-1} =
N^{1/D}>1$, we find,
\begin{equation}
\matrix{
\frac{M_L}{M_0} =  x^{DL}, \qquad
\frac{V_L}{V_0} =  x^{3L}, \qquad
\frac{U_L/M_L}{U_0/M_0} = \frac{x^{(D+p)(L+1)}-1}{x^{(D+p)}-1}, 
 \hfill \cr
\frac{T_L /V_L}{T_0/V_0} =  1 -\beta\Bigg[
\frac{1}{1-x^{2D+p-3}} + \frac { x^{L(D-3)}}{ x^{D+p} -1 } 
\left( 1 - x^{(L+1)(D+p)} \frac{1- x^{D-3}}{1 - x^{2D+p-3}}
 \right) \Bigg] \hfill \cr
}
\end{equation}
where $\beta \equiv pU_o/2T_0$ is the ground virial ratio ($0\leq
\beta \leq1$).  The free parameters are $D$, $N$, $p$, and $\beta$.
Although not very illuminating at first sight, the functional
properties of the solution (7) are very interesting.  We just
summarise here the most important features.  More detail will be
presented elsewhere (Pfenniger 1996).

\def\half{{\textstyle\frac{1}{2}}}

First, not any combinations of parameters lead to a physical solution.
Not only the kinetic energy $T_L$ must positive, but also the
pressure, which is proportional to the velocity dispersion $v_L$
squared, $\half M_L v^2_L = T_L - N T_{L-1}$.

In the ``thermodynamical limit'' $L\to \infty$ a striking phenomenon
occurs. The range of physical solutions shrinks on a subspace of the
parameter space, leading to a new constraint,
\begin{equation}
D =\frac{3-p}{2-\ln(1-\beta)/\ln N}.
\end{equation}
So, for $p=-1$, $D<2$ in any case. For systems not too confined by the
outer pressure ($1/2<\beta<1$) and $N>5$, we have $D < 1.7$, while
systems with more outer pressure (``pressure confined clouds'')
increase $D$ up to 2.  {\it Therefore, we conclude that hierarchical
gravitating systems in statistical equilibrium can indeed exist, but
with $D<2$.}

The scale-velocity dispersion relation takes the exact scale-free form
$v \propto r^{\kappa}$, where $\kappa\equiv (D+p)/2$. The result for
$p=-1$ in PC94 remains therefore unchanged in spite of the inclusion
of the outer pressure.  For fractal ISM cold clouds with substantial
ambient pressure (Blitz 1993), $\beta \approx 1/2$, and with $N=5-10$
(Scalo 1990), we expect $D \approx 1.7$ and $\kappa \approx 0.35$,
which is comparable to Larson's (1981) size-linewidth relationship.

Finally, the short to long-range interaction transition occurs at
$p=-3$.  For $p<-3$ the typical feasible state is non-longer
hierarchical, but with a single level, large outer pressure ($\beta
\ll 1/2$), large $N$ and a dimension $D$ close to 3: the classical
homogeneous gas is recovered.

\section{Final Remarks: Universality of the Fractal Structure}
With a simple model we have motivated here the proposition that
strongly gravitating systems in statistical average tend to lower
their dimensionality below 2, so to adopt a very inhomogeneous
structure, as manifest in many cases. The property $D\lta 2$ is well
documented for the cold ISM (Scalo 1990), and for the large scale
structures (Coleman \& Pietronero 1992).  Actually, the general
transparency, or blackness, of the Universe at most of the
wavelengths, extending the De Ch\'eseaux-Olbers paradox to non-stellar
objects, reflects for a good part the widespread $D<2$ fractality.

A deceptive effect occurs when observing fractals with $D<2$, such as
cold ISM clouds: their orthogonal projections have the same $D<2$
(Falconer 1990), so less than a surface.  Practically it means that
over the more scales the gas does indeed behave as a fractal, the
smaller is the sky fraction over which 50\% of the mass projects.  The
bias is then obvious: when sampling the sky, most of it appears at low
column density so mass {\it looks\/} well sampled.  In fact the mass
is poorly sampled because most of it resides in small regions of the
sky at high column density, therefore likely to be optically thick.
Not only increasingly higher angular resolution is required to resolve
the smallest structures, but also more wavelength types for piercing
all the decades of column densities.  In addition, the velocity
dispersion (or temperature) decreases at smaller scales, so the smallest
structures are both the coldest and the densest ones, so often hard to
detect.

If the fractal state in fast cooling cosmic gas is indeed universal,
we have little reason not to expect that the outer HI galactic disks
are just the warmer ``atmosphere'' of a fractal which extends down to
very small scale.  The lowest temperature in the Universe being 2.73
K, it is then natural to assume that much more gas mass can be hidden
in very cold molecular gas, making a sizable fraction of the galactic
dark matter (PC94).

\end{document}